# TripHLApan: predicting HLA molecules binding peptides based on triple coding matrix and transfer learning

Meng Wang[1], Chuqi Lei[1], Jianxin Wang[1], Yaohang Li[2], and Min Li[1,*]

[1] School of Computer Science and engineering, Central South University, Changsha 410083, China.

[2] Department of Computer Science, Old Dominion University, Norfolk, VA 23529, USA.

*To whom correspondence should be addressed.

## Abstract

Human leukocyte antigen (HLA) is an important molecule family in the field of human immunity, which recognizes foreign threats and triggers immune responses by presenting peptides to T cells. In recent years, the synthesis of tumor vaccines to induce specific immune responses has become the forefront of cancer treatment. Computationally modeling the binding patterns between peptide and HLA can greatly accelerate the development of tumor vaccines. However, most of the prediction methods' performance is very limited and they cannot fully take advantage of the analysis of existing biological knowledge as the basis of modeling. In this paper, we propose TripHLApan, a novel pan-specific prediction model, for HLA molecular peptide binding prediction. TripHLApan exhibits powerful prediction ability by integrating triple coding matrix, BiGRU + Attention models, and transfer learning strategy. The comprehensive evaluations demonstrate the effectiveness of TripHLApan in predicting HLA-I and HLA-II peptide binding in different test environments. The predictive power of HLA-I is further demonstrated in the latest data set. In addition, we show that TripHLApan has strong binding reconstitution ability in the samples of a melanoma patient. In conclusion, TripHLApan is a powerful tool for predicting the binding of HLA-I and HLA-II molecular peptides for the synthesis of tumor vaccines.

**Contact:** Min Li at limin@mail.csu.edu.cn.

**Supplementary information:** https://github.com/CSUBioGroup/TripHLApan.git.

Human leukocyte antigen (HLA) is the most polymorphic gene group in the human population, with more than 16,200 different types [1, 2]. These groups of genes help detect and identify foreign threats and trigger immune responses. Immunotherapy has emerged as a powerful cancer treatment in recent years, accurately predicting whether a peptide can be presented by a particular HLA allele or not has become a critical problem in clinical immunotherapy. [3-5]. HLA molecules are mainly divided into two categories: Class-I and Class-II HLA. HLA Class-I molecules mainly present intracellular endogenous peptides to the surface of CD8+ T cells, which are mostly derived from the degradation of intracellular proteins. HLA Class-II molecules present exogenous peptides mainly through the phagocytosis of some phagocytes [6].

In the past 20 years, some computational tools have been developed to predict the binding of HLA molecules to peptides. For the prediction task of HLA-I peptide binding, early prediction tools, such as SYFPEITHI [20] and SMMPMBEC [21], mainly use probabilistic methods to calculate position specificity to establish models. In recent years, more forecasting tools adopt machine learning, particularly deep learning methods, to build models. These models[14, 15] have shown strong predictive power in this field. For example, netMHCstabpan [22] combines the immunogenicity characteristics of HLA-I-peptide complex to establish the artificial neural network model. ACME [23] uses the combination of CNN (convolutional neural network) and Attention modules to establish an interpretable model for affinity prediction. MHCSeqNet [24] applies Embedding [25] and skip-gram [26] models in NLP (Natural Language Processing) to HLA-I-peptide prediction. These models all show good predictive performance in specific environments. Similar to the development of HLA-I tools, the prediction tools of HLA-II molecule are gradually evolving from classical machine learning to deep learning. However, there are few tools for predicting HLA-II molecule binding to peptides, and their prediction accuracy are not high. At present, some successful tools, including NN-align[27], NetMHCIIpan[19], PUFFIN[28, 29], MARIA[30], MHCAttnNet[16], etc., only achieve good predictions on certain specific alleles[29, 31]. There are also some tools that can simultaneously predict the binding of two classes of HLA molecule with peptides, such as MHCAttnNet[16] and MHCnuggets[8], but they show poor predictive performance.

The current tools can be classified into allele-specific tools and pan-specific tools. The allele-specific tools train a unique predictive model for each class of HLA molecular allele type. For example, NetMHC 4.0 [7] uses a feed forward neural network to classify HLA-I molecules. MHCnuggets [8] firstly clusters alleles, trains a model for the allele with the most binding peptides, and then conducts transfer learning in clusters. Allele-specific tools often provide strong predictive performance, depending on the available amount of training data per allele and the quality of allele clustering [9, 10]. However, these tools have difficulty to generalize over all alleles, especially those that have not yet been discovered. Pan-specific approach [11-17] mixes all HLA data together to train a pan-specific model that can predicts all alleles with known sequences, such as MixMHCp [18] and the latest version of NetMHCpan 4.1 [19]. As more and more HLA alleles are discovered, pan-specific tools play an increasingly important role in predicting the binding of HLA molecules with peptides.

Although current tools achieve some predictive performance, they still have some fatal limitations. (1) The predictive accuracy of current tools is inadequate. Especially for HLA-II-peptides binding prediction task, only limited types of alleles that can be predicted, while the accuracy is not sufficiently high for practical applications. (2) Although the prediction performance of the existing prediction tools for certain peptide lengths with a lot of available samples is acceptable, once the training samples for longer peptide lengths become insufficient, the predictor performance decreases sharply. For HLA-I-peptides binding prediction task, current tools perform well on predicting peptides of lengths 9 and 10, have difficulty to extend to longer peptides. (3) Current prediction tools rarely use the biological properties and the connection of data known in the field. The characteristics of the data themselves may be masked in the data processing. In addition, most tools ignore the contextual information between amino acids. To address the above issues, we propose TripHLApan, a new pan-specific model, for predicting HLA-peptide binding. First, we have analyzed the small amount of 3D structure information of HLA-peptide binding in Uniprot database[33] and the statistics of the data, and have conducted a more appropriate data preprocessing progress for HLA molecules and peptides. Next, we construct comprehensive feature profiles of peptides and HLA molecules, including the physicochemical and biochemical properties of amino acids, the probability of substitutions between amino acids, and the endogenous hidden information related to binding. We design a new triple-channel BiGRU [34] + Attention [35] model to learn the potential information of peptide and HLA molecular sequence. The triple-channel model enables TripHLApan to accurately predict the binding relationship between peptides and HLA molecules. Performance comparisons on several test sets show that TripHLApan outperforms the current best model.

## Results

**Overview of TripHLApan.** TripHLApan is the first HLA-peptide binding prediction tool to encode and train in parallel using multiple amino acid characteristics (Figure 1). In this model, the peptide sequence and HLA sequence are preprocessed and encoded. Three different encoding matrices are used to represent the sequences. Then, the encoded sequences are fed into the BiGRU + Attention model, and the outputs are concatenated together to form a matrix. After that, the output matrix is passed into three fully connected layers and one sigmoid layer, where the final binding probability is computed. We carry out a 5-fold cross-validation to avoid the contingency of model training data partitioning. Each part of the model is described in detail in the Methods section.

One can find that TripHLApan's BiGRU + Attention model is well adapted to the binding problem between HLA molecules and peptides. Firstly, the combination of peptides and proteins is largely determined by the complementary nature of their 3D structures. In the case of 3D structure information missing, BiGRU model based on sequence context information can well capture such sequence-based global information instead. Secondly, after exported from BiGRU, TripHLApan does not directly take the last layer of the output matrix as the feature of the whole sequence like most models. Instead, it first uses the Attention model to redistribute the weights according to the importance of the sub-sequences learned by BiGRU model at the positions of various amino acid residues. Then it takes the last hidden layer as features of the whole sequence and inputs them into the next fully connected layer. The advantage of BiGRU + Attention method is that it is able to identify the influence of the subsequences at both ends of the peptide sequence, determining whether the whole peptide can bind or not. This is consistent with the phenomenon that the two ends of a peptide often show peptide-binding motifs. Studies [36-38] suggest that the binding process of peptides with length 8 to HLA-I may involve the change of HLA-I molecular structure. Therefore, in the process of HLA-I model training, we first use the data sets of peptides in length 9-14 for model training, and then transfer the model to peptides in length 8. This training strategy ensures that the binding model of peptide with lengths 9-14 is not influenced by the data of peptide with length 8, and also ensures that the prediction model of peptide with length 8 retains the learned HLA-I binding characteristics of peptides with lengths 9-14.

In order to better simulate the binding relationship between HLA molecules and peptides, we have analyzed the location characteristics of HLA binding peptides on different locus by Uniprot database. As shown in Figure 1c, protein 1AKJ is the binding structure of HLA-A and peptide. The peptide is completely wrapped into two helical structures and one β structure of HLA-A. Locating the key positions of several structures in the sequence index, we find that the positions of HLA-A binding peptide are all in the first 200 amino acids of the sequence. Several other class-I HLA molecules also show similar rules. Therefore, the first 200 amino acid fragments of HLA-I are selected to represent the whole HLA-I molecule and input into the model. For HLA-II, different scenes are observed. In Figure 1c, 1JK8 and 3LQZ respectively represent the binding relationship between HLA-DQA/DQB and HLA-DP1A/DP1B with peptides. The two chains of HLA molecule play a role together in the binding relationship with peptide. We specifically extract two 100mer subsequences of different locus according to their different peptide binding sites to represent the two chains of HLA-II molecules, and feed them into the model respectively. See Methods for more details.

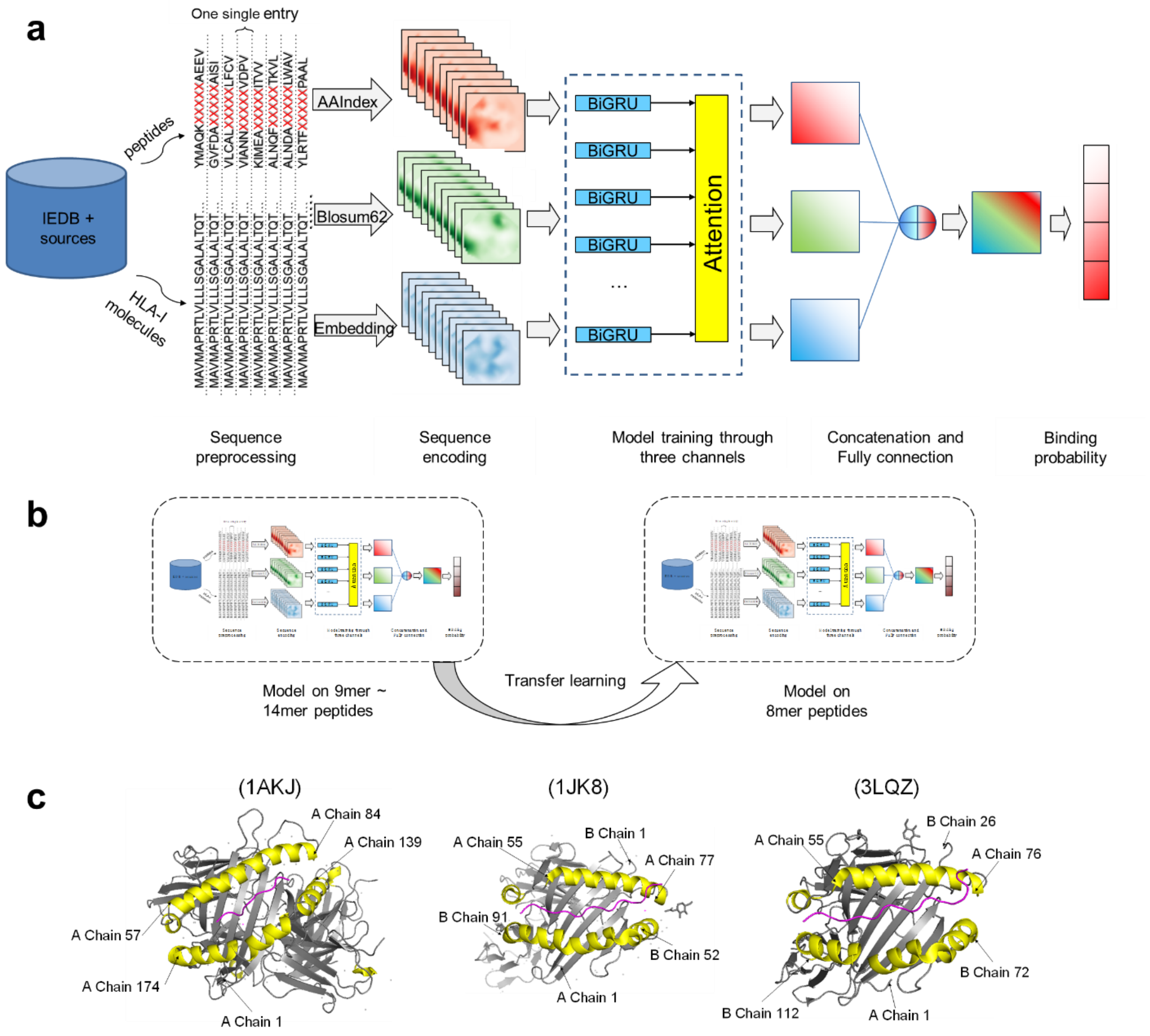

**Figure 1.** The workflow and architecture of TripHLApan. **a** Network architecture of TripHLApan for HLA-peptide prediction. TripHLApan firstly preprocesses HLA and peptide sequences with three coding matrices, then puts the coded matrices into BiGRU + Attention modules. After a fully connected layer at the end of each channel, the outputs of the three matrices are concatenated. Finally, after three fully connected layers, TripHLApan outputs the predicted binding probability from the sigmoid layer. **b** Transfer learning patterns on HLA-I-peptide prediction task. A model is trained on binding peptide lengths of 9-14 and is then transferred to the model with peptide length of 8. **c** Different binding forms of HLA-I and II molecules to peptides. 1AKJ: a complex of HLA-A molecules and peptides. 1JK8: a complex of HLA-DQA/DQB molecules and peptides.

3LQZ: a complex of HLA-DP1A/DP1B molecules and peptides.

**TripHLApan outperforms baseline methods on HLA-I binding prediction.** We compare TripHLApan with several carefully selected tools for classification performance, including the highly cited PickPocket [39], netMHCstabpan [22], and MixMHCp [18], which are considered the best predictive tool in a recent survey [40], and five recent pan-specific tools: MHCSeqNet [24], MHCflurry 2.0 [41, 42], and NetMHCpan 4.1 [19], MATHLA [17], MHCnuggets [8]. The test data set used to compare the performance of each tool is from the test set from publications and random negative peptide library (these negative peptides are obtained by random clipping from proteins, and the negative peptides used to fill the test and training sets is from two completely independent random negative peptide libraries), which is fair to the benchmarking tools. Even this testing is slightly unfair to TripHLApan on the unseen set since the benchmarking tool's training process contains more items from our unseen sets (see Figure 2a). The area under the receiver operating characteristics curve (AUC) is adopted to measure the classification ability of positive and negative samples of the prediction model. In order to eliminate the deviation of AUC values caused by the imbalance of positive and negative samples, AUPR is also used to compare their predictive performance. In addition, top-PPV (positive predict value) is estimated from the fraction of positive peptides within the top N predictions, where N denotes the number of true positive samples from the test samples. In the independent test set, we further define the test and unseen sets here. Their meanings are as follows: The allele types in the test set occur in the training set, and the allele types in the unseen set do not occur in the training set.

The results in Table 1 show that TripHLApan is a powerful predictive tool on both test and unseen sets. TripHLApan has not only strong predictive power but also strong generalization ability. Figure 2b shows that in the test set, TripHLApan outperforms all other tools in predicting peptides of different lengths on the rate of positive and negative samples at 1:5/1:1/1:10/1:50. Moreover, TripHLApan's performance degrades much slower than the other methods for longer peptides. In particular, on the rate of positive and negative samples at 1:5, TripHLApan's AUC value outperforms the second-ranked models by 1.2%, 6.1%, 18.8%, 10.3%, and 24.9% for peptide lengths of 10/11/12/13/14 in the independent test set, respectively. The unseen set, whose goal is to measure the model's generalization ability, has alleles not seen in the TripHLApan's training set. The results show that the AUC curves of TripHLApan are similar to those of the tools in comparison. Especially when the peptide length is over 12, TripHLApan exhibits significantly higher AUC values than the other tools. Furthermore, TripHLApan has the highest AUPR values (see Table 1 and Figure 2c), indicating that TripHLApan is able to handle different data set environments and can actually achieve good prediction results. Figure 2 c&d shows the AUPRs and top-PPVs of selected tools with positive and negative samples at 1:1/1:5/1:10/1:50. One can find that the prediction accuracy of TripHLApan is comparable to NetMHCpan, MixMHCp and MHCflurry on the unseen set, and it shows better prediction ability in other test sets. This suggests that TripHLApan can effectively learn the patterns between sequences and binding patterns and thus pick out the most likely combinations of potential compounds, even when the sample ratios are much more stringent, such as 1:10 and 1:50. Then, TripHLApan's strong learning ability on the new alleles is verified. We randomly select 4/5 portion of the unseen set to perform transfer learning using TripHLApan, and the remaining 1/5 portion is used for testing. The results are shown in Figure 2b 'unseen set (left)' where the prediction from TripHLApan is improved after transfer learning. This indicates that TripHLApan can adapt easily to a variety of prediction environments. Even on the allele conjugate that has never been seen before, TripHLApan can take advantage of known data for transfer learning to enhance its prediction capability. To eliminate the unseen set's unfairness to TripHLApan and the benchmark tools, we extract the alleles that do not appear in all tool's training sets (named 'all unseen set'), a total of 19,160 items on 13 HLA alleles. TripHLApan obtains the highest AUC/AUPR and top-PPV scores (see Figure 3). Therefore, we believe that, with the increase of HLA molecular-peptide data, TripHLApan can also achieve good prediction results after simple transfer learning on new allele data set.

**Table 1.** AUCs and AUPRs of TripHLApan compared with baseline tools on test set and unseen set.

| Methods | test set | | unseen set | |
|---|---|---|---|---|
| | AUC | AUPR | AUC | AUPR |
| **TripHLApan** | **0.978** | **0.933** | **0.958** | **0.881** |
| netMHCpan | 0.905 | 0.839 | 0.949 | **0.883** |
| MHCSeqNet | 0.815 | 0.354 | 0.853 | 0.470 |
| MixMHCp | 0.896 | 0.779 | 0.941 | 0.843 |
| MHCflurry | 0.933 | 0.871 | 0.938 | 0.861 |
| netMHCstabpan | 0.888 | 0.740 | 0.922 | 0.780 |
| PickPocket | 0.877 | 0.716 | 0.923 | 0.770 |
| MATHLA | 0.891 | 0.704 | 0.920 | 0.741 |
| MHCnuggets | 0.854 | 0.544 | 0.759 | 0.534 |

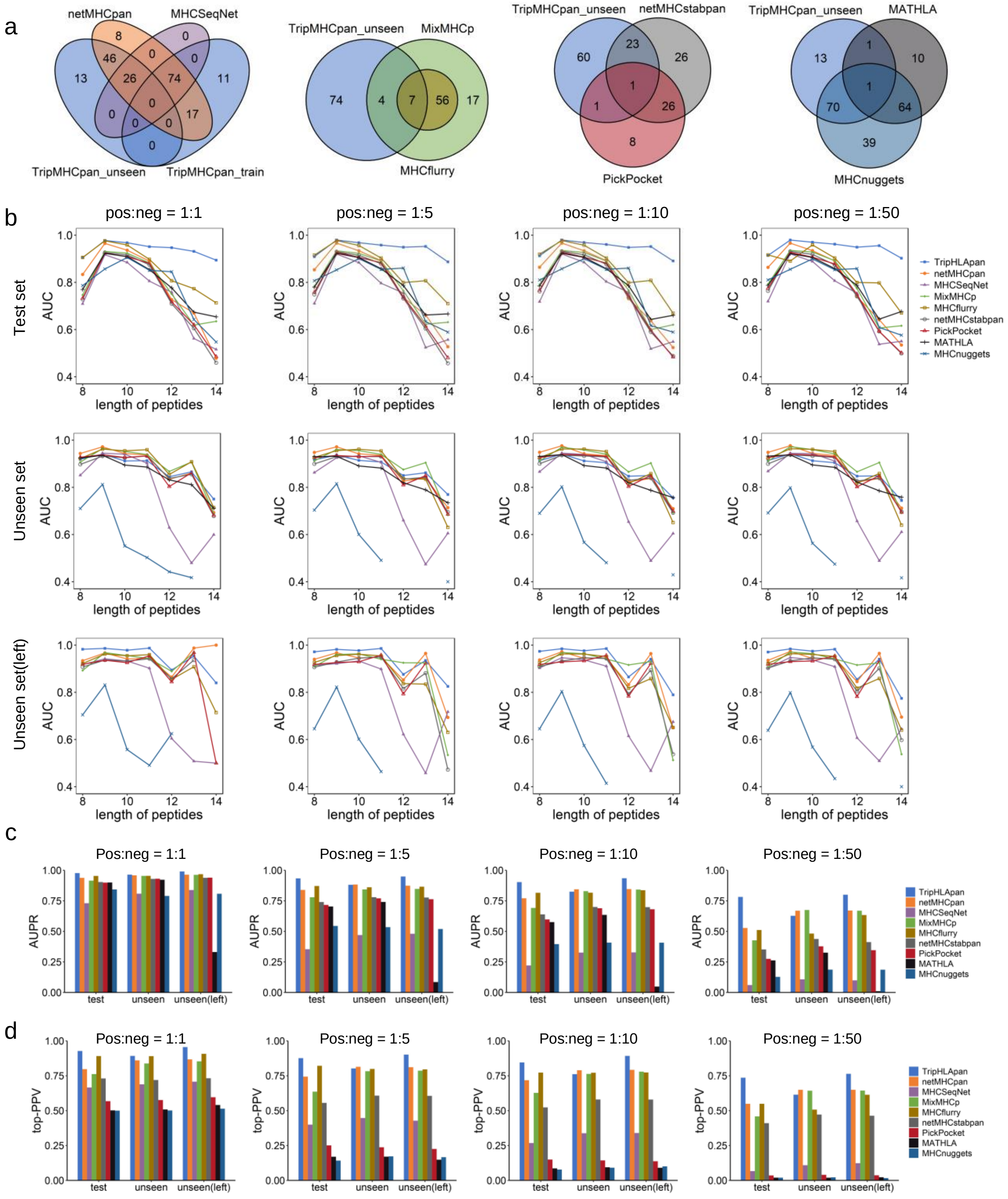

**Figure 2. TripHLApan is compared with the prediction results of baseline tools. a** The number of overlapping HLA types of the unseen set used in this paper and the training sets of 9 tools including TripHLApan. **b**/**c**/**d** AUCs/AUPRs and top-PPVs on the data sets with peptides of different lengths on the rate of positive and negative samples at 1:5/1:1/1:10/1:50.

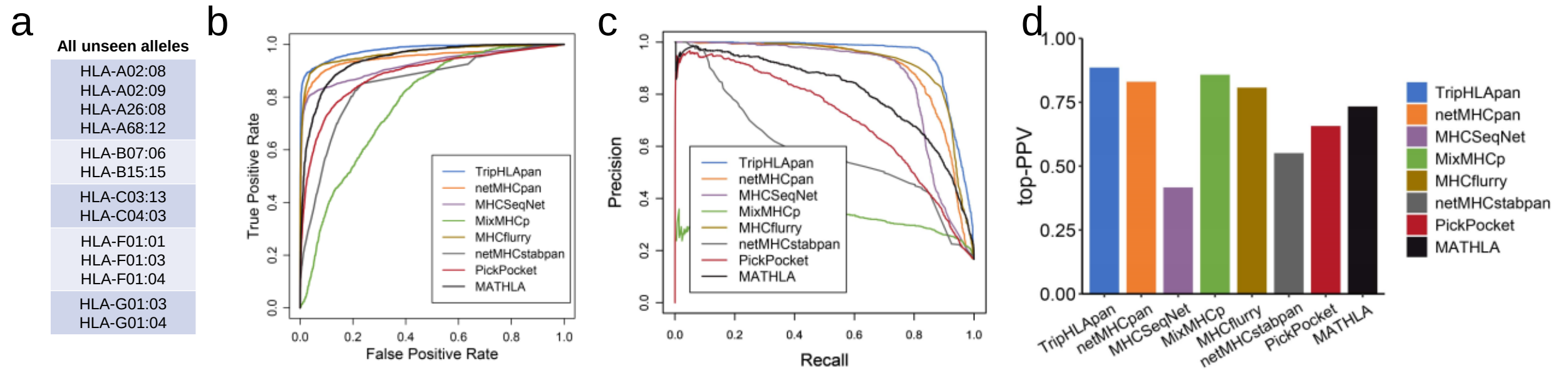

**Figure 3. TripHLApan is compared with the prediction results of baseline tools on the dataset in which the alleles do not appear in all tool's training sets. a** Allele types in the all unseen set. **b/c/d** ROC/AUPR curves and top-PPVs on the all unseen set.

**Testing on the latest data set.** We have tested TripHLApan on the recent data set from the IEDB database to further verify its validity. Firstly, we extract the recent data set from the IEDB (downloaded on October 7, 2021), and de-replicate it with the train set, test set, and unseen set used in this paper to ensure that the new test data are completely new to the model. Then, the new data set is purified in the same way described in this paper. Finally, a total of 68,277 data samples are obtained, including 14,705 positive samples. Figure 4 a/c/d shows that TripHLApan still achieves the best prediction performance in the recent test set compared with the three models that perform better in the test set. In addition, we divide the recent data set into different subsets based on different peptide lengths to evaluate the performance differences of the tool on different peptide lengths. After excluding data sets with sample numbers less than 50, the prediction results of different subsets are shown in Figure 4b. TripHLApan is also a leading method in predicting peptide groups of different lengths. In conclusion, we compare the prediction performance of different tools based on the latest epitope data collected by IEDB, and the results show that TripHLApan has a more stable prediction performance under different prediction environments.

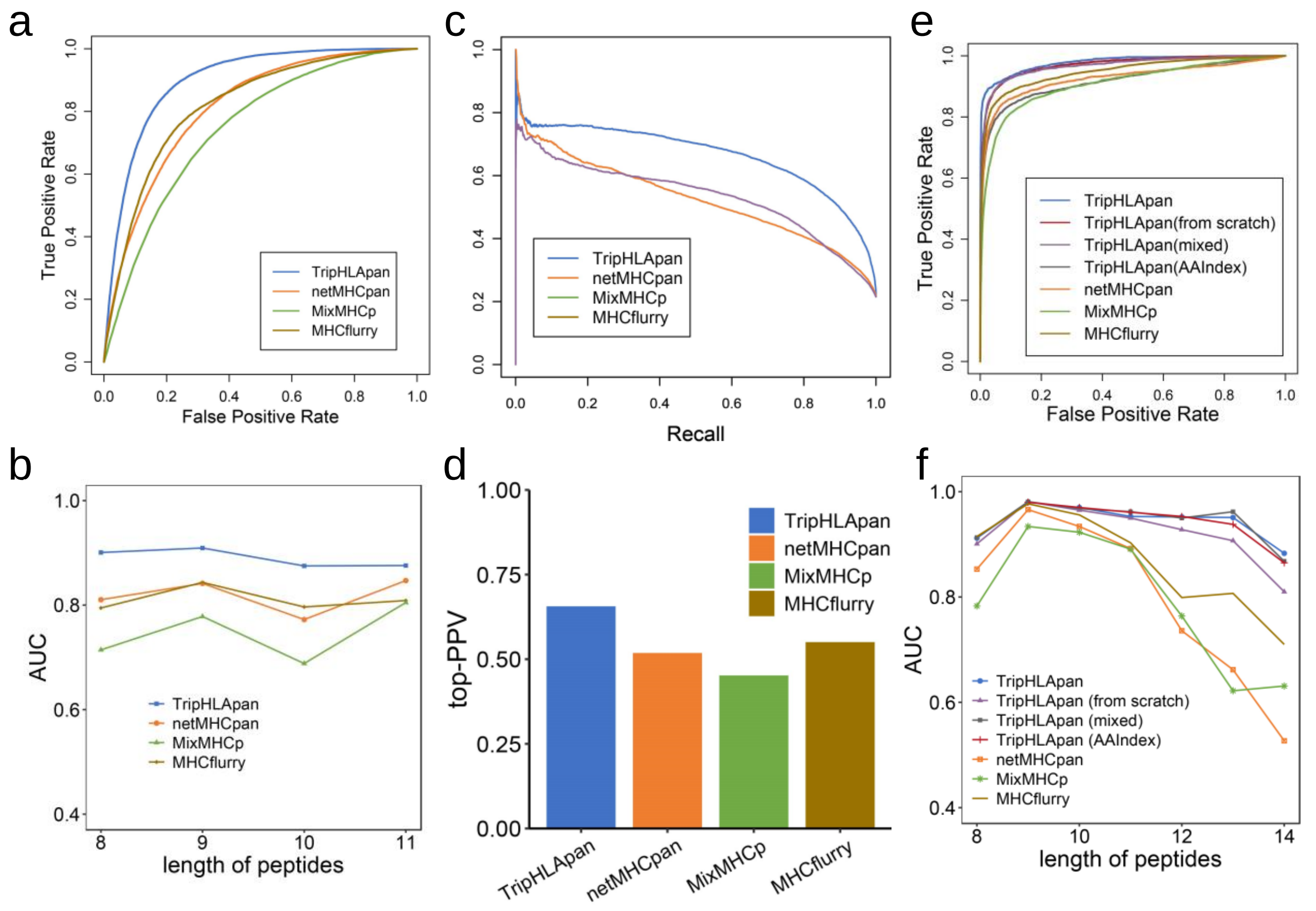

**Figure 4. Prediction results comparison of TripHLApan with baseline tools on the lasted data set and its ablation experiment. a/b/c/d** ROC/AUCs on peptides of different lengths/AUPR curves and top-PPVs on peptides of different lengths on test set of TripHLApan. **e&f** ROC curves and AUCs on peptides of different lengths on the test set of TripHLApan and its three different model strategies, as well as the three baseline models that performed better.

**Proper model architecture and data preprocessing contribute to better model performance.** We attribute TripHLApan's strong predictive power to its appropriate model combination and correct strategy selection. Firstly, three parallel sequence encoding methods are used to extract the information contained in the sequence from the perspectives of biochemical properties of amino acids, substitution probability and inherent hidden information related to binding. It is important to note that the biochemical properties of amino acids have been neglected by most methods, but these properties turn out to be crucial in predicting the combination of peptides and HLA molecules. Secondly, BiLSTM and BiGRU models, which can learn the relationship between sequence context information, are selected not only because they are as effective as described, but also because they do show strong predictive power in previous studies [24]. According to our preliminary comparative experiments, BiGRU is selected as our basic model unit. Thirdly, we add Self-Attention modules into TripHLApan to pay attention to the weight of each subsequence of peptides and HLA-I molecules. We choose this kind of Self-Attention approach because of the observed phenomenon of ‘significant anchor residues at both ends of the peptide’ [18]. Selecting such modules allows TripHLApan to better capture the complex correlations of the anchor sites and their adjacent sites, and maximize the exposure of anchor residues at both ends of the sequence. Finally, the transfer learning strategy reduces the performance loss caused by excessive mixing of data and retains the learned binding modes. In addition, the ‘dropout’ and ‘early termination’ strategies also play an important role in preventing overfitting and improving model stability.

Table 2 shows the results of some ablation models compared with the original model. One can find that TripHLApan is the model with the strongest comprehensive prediction ability. However, TripHLApan (AAIndex), which is encoded using only the biochemical properties of amino acids, also achieves very similar prediction results, indicating the biochemical properties of amino acids are the most crucial features. The training strategy of transfer learning also helps improve the performance of TripHLApan on the data sets with peptides of different lengths. Figure 4 e&f compares TripHLApan in detail with the model without transfer learning strategy [TripHLApan (mixed)], the model with ab initio training for different peptide lengths [TripHLApan (from scratch)], and the well-performing TripHLApan [AAIndex]. The training strategy of transfer learning limits the prediction ability of peptide length 8 to some extent, but it is extended for longer peptide length (11/12/13/14). By comparing the results in Table 1 and Figure 4 e&f one can find that TripHLApan’s prediction performance at different lengths is occasionally worse than that of some reference tools, but its overall AUC is much better than the other tools. This suggests that TripHLApan’s ability to separate positive and negative samples remains stable across data sets with peptides of different lengths. This is an important ability for predictive tools, especially when most current tools use defined thresholds for dividing positive and negative samples. In general, TripHLApan collects more comprehensive sequence information and amino acid characteristics, and adopts biological prior knowledge. As a result, it greatly improves the accuracy of binding prediction of HLA-I molecules with peptides.

**Table 2. AUCs of model ablation experiment.**

| Models | TripHLApan | 1# | 2# | 3# | 4# | 5# | 6# | 7# |
|---|---|---|---|---|---|---|---|---|
| test set | 0.981 | 0.980 | **0.985** | 0.978 | 0.978 | 0.979 | 0.977 | 0.979 |
| unseen set | 0.942 | 0.938 | 0.933 | 0.937 | 0.935 | **0.944** | 0.922 | 0.916 |

These abbreviations represent different versions of TripHLApan: 1#: without Self-Attention; 2#: AAIndex + Blosum62; 3#: AAIndex + Embedding; 4# Blosum62 + Embedding; 5#: AAIndex; 6#: Embedding; 7#: Blosum62.

**TripHLApan can predict Single-Patient Cancer Immunopeptidome.** To test the clinical application of TripHLApan, we collect individual patient melanoma-related Immunopeptidome data sets [43-45]. In this data set, the true binding relationship between HLA-I molecule and the peptide is unknown. We use TripHLApan to reconstruct the HLA-I molecule and peptide correspondence in various cell lines. TripHLApan's ability to re-annotate peptide presentation molecules is tested by calculating the Pearson correlation coefficients (PCCs) between the frequencies of amino acids in the predicted positive sample at each location and the frequencies of amino acids in the real binding. The comparison tools NetMHCpan and MixMHCp are excluded because they have already included these four cases in their training set. We also exclude PickPocket, MATHLA, and MHCnuggets here, due to their poor performance on the top-PPV according to Figure 2d. Figure 5 shows the average PCCs of tools on the data sets with peptide lengths at 9 and 10. TripHLApan has demonstrated a powerful ability to reconstruct peptide-HLA-I molecular relationships.

TripHLApan's deconvolution ability is more stable in various cell lines compared to other methods. Especially in Mel5 and Mel8, when the peptide length increases from 9 to 10, the relationship reconstruction ability of all other tools decreased significantly, while TripHLApan remains stable in contrast. This not only demonstrates TripHLApan's ability to reconstruct relationships with peptide samples of length 9 (the most common peptide length), but also shows TripHLApan's stability over longer peptide samples. For more detailed results, see Supplementary Table S1 and S2.

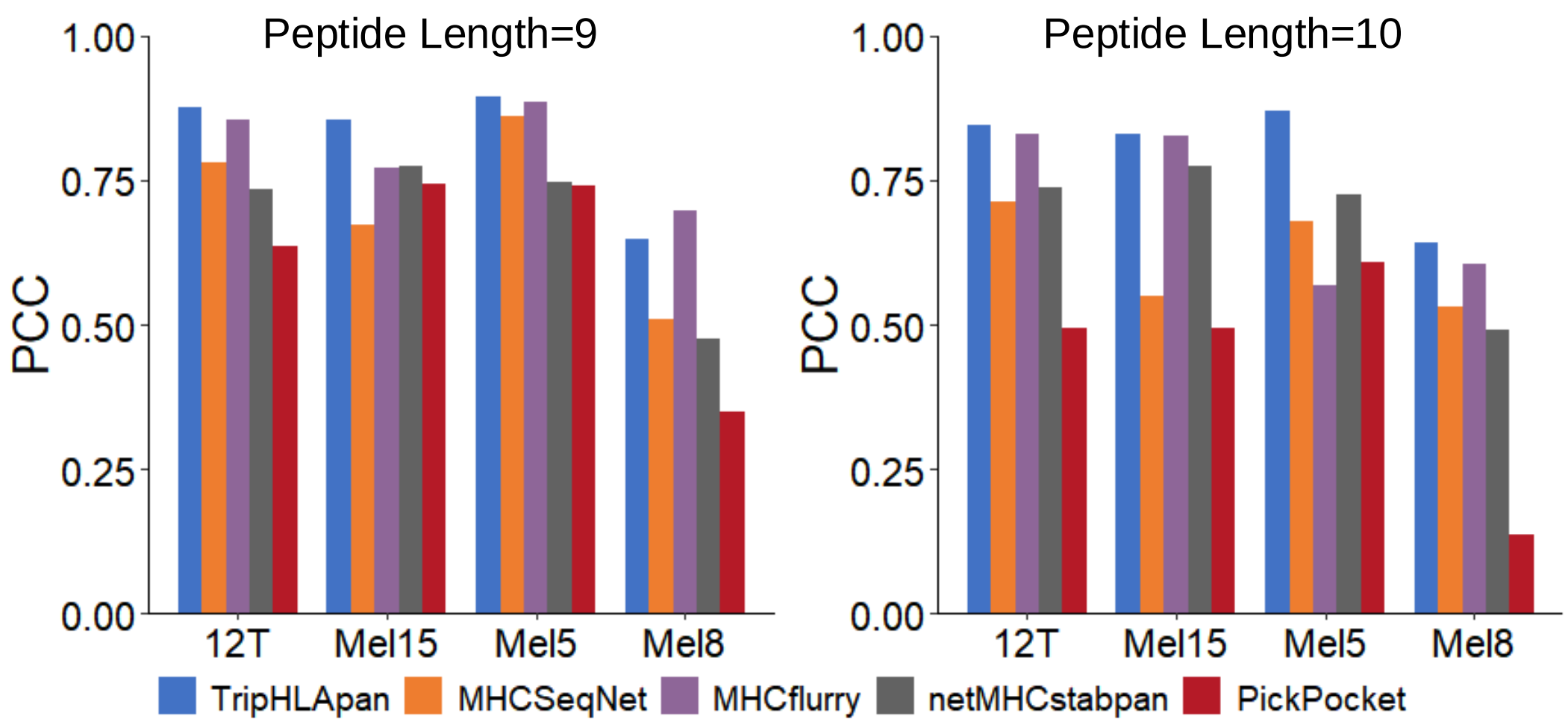

**Figure 5. Comparison of the average PCCs of tools on the four Mono-Allelic samples.**

**TripHLApan outperforms baseline methods on HLA-II binding prediction.** For the task of predicting the binding of HLA-II molecules to peptides, the current tools have the following limitations: (1) the molecular types that can be predicted are limited; and (2) the prediction accuracy is generally low. Based on the above analysis of the only 3D structure data of HLA-II molecules and the outstanding performance of TripHLApan in HLA-I, we find that TripHLApan also has strong predictive ability in predicting the binding of HLA-II molecules and peptides. Similar to predicting Type I, data in the independent test set is divided into the test sets and unseen sets according to whether its allele type appears in the training set or not. Finally, 77 test sets and 63 unseen sets are divided according to alleles. TripHLApan has AUCs greater than 0.9 on 61, 55 subsets of the test set and unseen set, respectively. In particular, TripHLApan's results on the 30 alleles in the unseen sets are above 0.99, an important breakthrough in the development of predictive models for HLA-II molecular peptide binding. We compare the results with 9 tools[27, 46-50] integrated into the IEDB database. Excluding DP and DQ that cannot be predicted by the baseline tools and other unsupported alleles, there are 35 subsets in the test set and 8 subsets in the unseen set. TripHLApan is able to achieve optimum AUC values on their 23 and 8 Alleles (Supplementary Table S3 and Table S4). Figure 6 shows the AUC distribution comparison results on the allele sets that most of the tools can predict, where TripHLApan (AAIndex) denotes the result of a version of TripHLApan only using AAIndex features. Next, we compare the performance of MHCnuggets in class II that can also do the binding prediction task of HLA-I and II peptides. The results show that TripHLApan can achieve better prediction accuracy on 31 alleles in 36 alleles collections supported by MHCnuggets. More detailed results are also shown in Supplementary Table S3 and Table S4. Overall, TripHLApan, as a pan-specific model that can predict more allelic types, has achieved a prediction accuracy not reachable before.

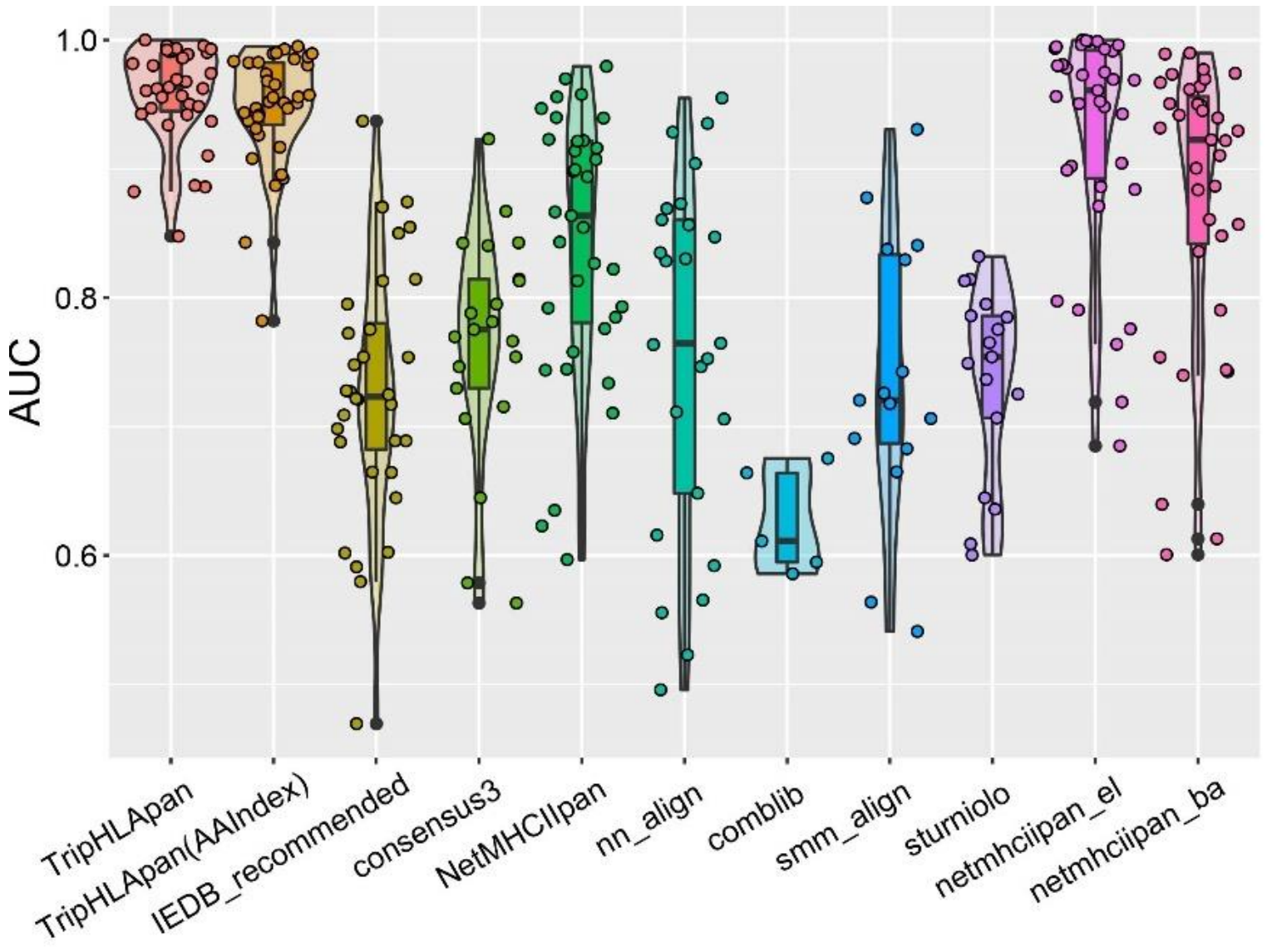

**Figure 6. AUC distribution of TripHLApan compared with baseline methods on 35 Allele sets on HLA-II-peptide binding prediction.**

## Discussion

In this work, we propose TripHLApan, a new predictive model of HLA-I binding with peptides integrating multiple features. We first analyze the biological and statistical characteristics of HLA molecule-peptide binding and obtain a more appropriate amino acid sequence preprocess scheme. Then, we integrate the biochemical characteristics of amino acids rarely used in other tools, and form a parallel sequence coding and training network with the probability of amino acid replacement and Embedding features. Finally, we use Attention module to learn the context information contained in each local sequence after BiGRU, avoiding the loss of sequence information caused by excessive attention to position-specific factors in the modeling process. TripHLApan is confirmed to be superior to the current state-of-the-art prediction tools by a 5-fold cross-validation. On the task of HLA-I-peptide binding prediction, we also test the generalization ability of the model with allele binding sample sets that do not appear in the TripHLApan training set (Note that this is unfair to TripHLApan, as these samples may have been used to train for baseline tools). TripHLApan's prediction results are still at the forefront of comparison tools. After a simple transfer learning in the unseen set, TripHLApan's results remain stable, demonstrating TripHLApan's powerful ability to learn and adapt to its environment. TripHLApan achieves the highest AUC score of 0.979 on the 13 HLA alleles that do not appear in all tools' training sets. Also, TripHLApan's powerful predictive performance has been validated on the latest data set. Nevertheless, we conduct a series of model ablation experiments, which confirm the effectiveness of the framework and strategy adopted in TripHLApan. Finally, we validate the stability of TripHLApan's predictive power in clinical use on a single patient melanoma-related Immunopeptidome dataset. In the HLA-II-peptide binding prediction task, we have achieved more significant improvement in prediction ability, which is mainly reflected in the following two aspects :(1) TripHLApan is able to predict more allele types. TripHLApan is a pan-specific prediction model that can be used to predict HLA molecules and peptides as long as their sequences are known. (2) TripHLApan has demonstrated effectiveness in HLA-II-peptide binding prediction task.

However, there are some limitations to TripHLApan. The enhancement is not quite observable on predicting the samples with the most common length of peptides at 9 on HLA-I-peptide binding prediction. This is mainly due to the fact that the predictive power on peptide length at 9 of existing tools has already reached a highly accurate point. In this case, the predictive power of the model will be more affected by the quality of the data. On the other hand, the characteristic of the TripHLApan's pan-specificity limits its performance for 9 lengths to some extent. In addition, the sequence-based dichotomous prediction models cannot determine the specific binding site and posture, and we will take 3D features into account in future studies, especially in this era when protein structure prediction has achieved fine-grained accuracy.

## Methods

**Datasets.** The data sets of HLA-I-peptide used for training and testing in this paper are from the IEDB database (downloaded on March 15, 2021) [50, 51] and several publications [8, 19, 23, 45, 52]. The data sets of HLA-II-peptide are from the IEDB database (downloaded on March 15, 2021). Only the items with mass spectrometry eluted ligands (in short, MS EL) label are selected for model training. Compared with affinity label, MS EL label covers antigen processing and presentation steps, and its experimental conditions are more standardized. The HLA allele sequences are typically collected from the Immuno Polymorphism Database [2].

The purification process of data sets is described as follows: (1) deleting the unlabeled items from all sources, and merging them into three categories: items from the IEDB database, items from train set, and items from test set referred to publications; (2) appending data from the IEDB database to train set; (3) deleting the items that appear in train set from the test set; and (4) de-duplicating the train set and test set separately. The repeated items' final labels are decided by majority vote. If the votes are tied, these items are removed. In this paper, alleles with more than 1,000/50 (1,000 for HLA-I, 50 for HLA-II) binding samples are divided into the train/test sets according to the original train/test sets, and the remaining allele items are treated as the items of the unseen set. The difference between the unseen set and the test set is that, the peptide sequences in training and test sets are completely separated, while in the unseen set, the peptides and HLA molecules are both "unseen" in the training set. 2 million sequences with length 8-32mer are cut from Uniport database [33] as random negative sample library. The final training/test/unseen set is obtained by a 5-fold negative sample filling. If the number of negative samples in the original negative sample set is less than 5 times the number of positive samples, the peptides will be randomly obtained from the random negative sample library as the negative samples. For HLA-I, the number of samples in the final training set is 2,788,602 (positive sample number: 464,767); the number of samples in the test set is 21,246 (positive sample number: 3,541); and the number of samples in the unseen set (referring to the data set of HLA-I molecules not occurred in the training set, used to measure the generalization ability of the model) is 66,096 (positive sample number: 11,016). For HLA-II, the number of samples are 963,186 (positive: 160,531), 192,846 (positive: 32,141), and 5,262 (positive: 877), respectively.

**Preprocessing and data encoding.** In preprocessing HLA-I molecular sequence, most of the previous prediction tools use the extraction of fixed position residues as the form of pseudo-sequences. But pseudo-sequences cannot capture the contextual relationship between adjacent amino acids within the sequence, which may lose important biological information. In order to find out the effect of encoding methods on the results, we investigate the relationship between sequence similarity and motif similarity of binding peptides under several different encoding schemes. We select 45 alleles with positive samples of more than 100 peptides and calculate the similarity between their sequences and the similarity of peptide motifs respectively. The results are illustrated in Figure 7. There are 11 allele families with similar binding peptide motifs. Pair similarity based on full sequence of alleles can capture the similarity of 10 motif groups except the first one, while the other two pseudo sequences [38, 53] can capture only 7 motif groups. This indicates that these two pseudo sequences can only capture part of the key information of alleles binding to the peptides, but they have difficulty in extracting more comprehensive information. Allele representation based on the full sequence is a more stable and comprehensive way for sequence information extraction. The allelic similarity based on the full sequence is poorly matched with the binding motifs of the first group, which may be due to the low similarity of the parts of the sequence unrelated to the binding relationship in this group of alleles. This suggests that extracting the useful sequence information embedded in the full sequence is useful for modeling their binding process.

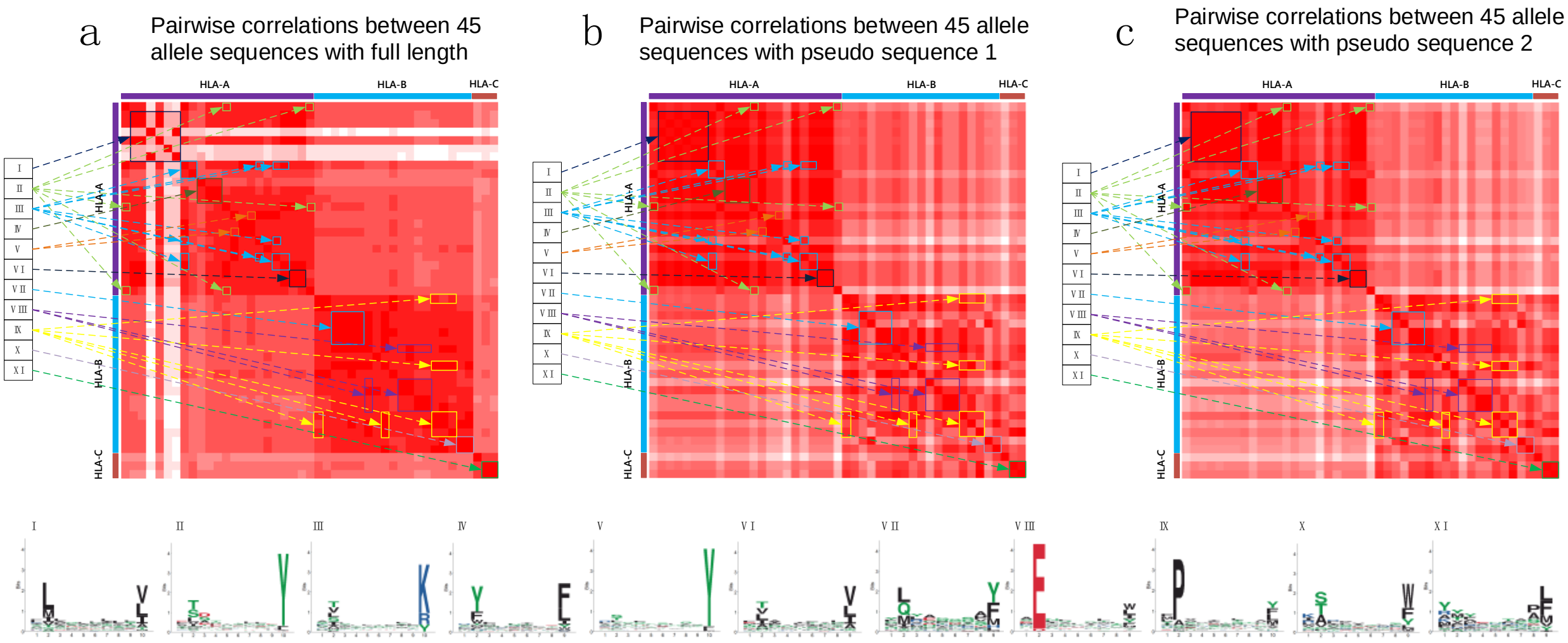

**Figure 7**. Identification of shared motifs amongst HLA-A, B and C alleles. Subfigures a, b and c illustrate the pairwise correlations between 45 allele sequences obtained by three allele sequence extraction methods and their corresponding relationships with motifs respectively. I to XI represent different motif groups.

According to the above conclusion, we adopt the full sequence of HLA molecules as the input of TripHLApan. For HLA-I, if the length of HLA molecules sequence is greater than 200, taking the first 200 amino acids from its sequence as input; otherwise, padding 'X's to length 200, indicating empty spots. For HLA-II, different amino acid fragments are extracted according to UniProt structural information. The starting positions of the different bases are 26 for DRA, 30 for DRB, 24 for DQA, 33 for DQB, 27 for DMA, 19 for DMB, 29 for DPA, and 30 for DPB. The undersized HLA molecules will follow the same pattern as the class I padding with 'X's. In peptide sequence preprocessing, according to the previous studies and our analysis above, we add 'X's in the middle of peptide sequences to the length of 14/32. This sequence processing strategy not only maintains the binding peptide motif of HLA molecules but also retains the context information of peptide sequence.

After preprocessing the HLA-I molecular sequences and peptide sequences, they are formatted into sequences of length 200 and 14, respectively. The lengths of these sequences for HLA-II molecular are 100, 100 and 14. In order to represent the information contained in the sequence, we use the following three different coding matrices to encode the sequences: (1) AAIndex[54]: AAindex is a database of numerical indicators representing various physicochemical and biochemical properties of amino acids and amino acid pairs. We select 28 features that have been successfully applied in other fields and are relatively consistent with our prediction task as an encoding of physical and chemical features[55]. (2) Blosum62 [32]: the Blosum62 matrix represents the probability of substitutions between pairs of amino acids; and (3) Embedding [25]: Embedding is a learnable encoding method that converts discrete variables into continuous number vectors. It continuously updates the values in these vectors through the back propagation of model training, to optimize the encoding of the sequence.

**BiGRU module.** Recurrent Neural Network (RNN) is a kind of deep learning model that can process time series information, which has been widely used in natural language and biological information processing. Long short-term memory (LSTM) and Gate Recurrent Unit (GRU) are the two most common variants of RNN, they are designed to address the common problems of gradient disappearance and gradient explosion in RNN. Compared to LSTM, GRU is a simplified version, showing better performance in many tasks [34]. The GRU updates the current output by adjusting the weights of the update gate, reset gate, and current state. BiGRU is composed of unidirectional and opposite GRUs whose output is determined by the states of the two GRUs. The main calculation formulas are:

$$z_t = \sigma(W_z \cdot [h_{t-1}, x_t]),$$

where $z_t$ represents the value of the update gate, $\sigma$ is the sigmoid activation function, $W_z$ is the weight matrix, $h_{t-1}$ is the state at the last moment, $x_t$ represents the input of this time, and [] denotes the concatenating of two matrices.

$$r_t = \sigma(W_r \cdot [h_{t-1}, x_t]),$$

where $r_t$ represents the value of the reset gate and $W_r$ is the weight matrix.

$$\widetilde{h_t} = tanh(W \cdot [r_t * h_{t-1}, x_t]),$$

where $\widetilde{h_t}$ represents the input gate of the current state, tanh(.) is the activation function, and $W$ represents the weight of the input gate.

$$h_t = (1 - z_t) * h_{t-1} + z_t * \widetilde{h_t},$$

where $h_t$ is the updated state at the current point in time, a balance between the current input and the past state.

**Attention module.** The Attention module [35] serves the purpose of focusing the key information on the sequence by redistributing the weight information in the sequence. The introduction of the Attention module complements important residues in the learning sequence. Attention module can be divided into different types according to the inputs. Here, we adopt Self-Attention, and the calculation formula is as follows: $Attention(Q, K, V) = softmax(\frac{QK^T}{\sqrt{d_k}})V,$ where $Q, K, V$ represent the queries, keys, and values matrices, respectively. In this paper, the three matrices are the same for the same encoding of the input. The attention matrices of different coding schemes are trained separately. $d_k$ is the sum of the values in each row of $QK^T$.

**Transfer learning.** TripHLApan uses transfer learning strategies in HLA-I-peptide prediction task to adapt to the different data of the model. First, a prediction model is trained from the data of peptide lengths 9-14 (taking the prior knowledge that molecular structure changes may also occur in the binding with peptide length 8 into account). Then, the trained model is transferred to the training set of peptide length 8, as shown in Figure 1b. By adopting this transfer learning strategy, TripHLApan can eliminate noises caused by data jumbling to certain extent. In addition, known binding information is used to pre-train the model for peptides of length 8, and the binding characteristics learned are retained. This training mode can effectively improve the accuracy of the model, whose details are discussed in the Result Comparison section.

**Network Training.** Figure 1 shows TripHLApan's workflow for HLA-I-peptide binding prediction task. HLA sequence and peptide sequence are encoded into a 200×28 matrix and a 14×28 matrix by AAIndex, a 200×20 matrix and a 14×20 matrix by Blosum62, and a 200×6 matrix and a 14×6 matrix by Embedding. Therefore, through data encoding, a total of 6 encoded matrices are obtained. Then, these six matrices are sent to six independent BiGRU networks, each containing 128 hidden layers to learn the relationship in the direction of each sequence. After the BiGRU layer, the size of the six matrices becomes 200×256, 14×256, 200×256, 14×256, 200×256, and 14×256. Generally, models only take the vector of the last dimension as the output of the next layer after passing through the GRU layers. Due to the particular binding relationship between HLA molecules and peptides, we use the Attention module to learn the importance of information between each consecutive hidden layer pair in BiGRU. Through BiGRU + Attention model, the machine learning model can learn the importance matrix of each sub-sequence. After going through the Attention model, six 256-dimensional vectors are obtained. HLA sequences and peptide sequences using the same encoding matrix are then concatenated together along the vector direction and pass through a 512×128 full connection layer. Then, the three 128-dimensional vectors are concatenated together along the vector direction. Next, the resulting vector will pass through a 384×128 fully connection layer, ReLu function, and a 128×128 fully connection layer, ReLu function. After removing 20% of the weight parameters through the 'dropout' function in PyTorch, the 128-dimensional vector passes into the last 128×1 full connection layer with a sigmoid activation function, which can be used to derive the predicted binding probability.

The learning rate is initialized to 0.0001. In addition, we adopt the scheme of decaying learning rate. The 'early termination' strategy is adopted to obtain the best trained model before overfitting. If the loss does not reduce in four consecutive iterations, the learning rate will be reduced by half.

HLA-II-peptide binding prediction models differ slightly from class I models in that the two chains of HLA are input separately in the first output step and then splice before full connection layer. See our Github repository (https://github.com/CSUBioGroup/TripHLApan.git) for more implementation details.

## Funding

This work is supported by the National Natural Science Foundation of China under Grant No. 61832019, the science and technology innovation program of Hunan Province (2021RC0048), and the Hunan Provincial Science and Technology Program (2019CB1007).

This work is also supported in part by the High Performance Computing Center of Central South University.

## References

1. IMGT, R. the international ImMunoGeneTics information system R 25 years on. *Nucl. Acids*.
2. Robinson, J. et al. The IPD and IMGT/HLA database: allele variant databases. *Nucleic acids research* **43**, D423-D431 (2015).
3. Siegel, R.L., Miller, K.D. & Jemal, A. Cancer statistics, 2019. *CA: a cancer journal for clinicians* **69**, 7-34 (2019).
4. Finck, A., Gill, S.I. & June, C.H. Cancer immunotherapy comes of age and looks for maturity. *Nature Communications* **11**, 1-4 (2020).
5. Gubin, M.M. & Schreiber, R.D. The odds of immunotherapy success. *Science* **350**, 158-159 (2015).
6. Neefjes, J., Jongsma, M.L., Paul, P. & Bakke, O. Towards a systems understanding of MHC class I and MHC class II antigen presentation. *Nature reviews immunology* **11**, 823-836 (2011).
7. Andreatta, M. & Nielsen, M. Gapped sequence alignment using artificial neural networks: application to the MHC class I system. *Bioinformatics* **32**, 511-517 (2016).
8. Shao, X.M. et al. High-throughput prediction of MHC class I and II neoantigens with MHCnuggets. *Cancer immunology research* **8**, 396-408 (2020).
9. Vang, Y.S. & Xie, X. HLA class I binding prediction via convolutional neural networks. *Bioinformatics* **33**, 2658-2665 (2017).
10. Bravi, B. et al. RBM-MHC: A Semi-Supervised Machine-Learning Method for Sample-Specific Prediction of Antigen Presentation by HLA-I Alleles. *Cell systems* **12**, 195-202. e199 (2021).
11. Karosiene, E., Lundegaard, C., Lund, O. & Nielsen, M. NetMHCcons: a consensus method for the major histocompatibility complex class I predictions. *Immunogenetics* **64**, 177-186 (2012).

12. Han, Y. & Kim, D. Deep convolutional neural networks for pan-specific peptide-MHC class I binding prediction. *BMC bioinformatics* **18**, 1-9 (2017).

13. Liu, G. et al. PSSMHCpan: a novel PSSM-based software for predicting class I peptide-HLA binding affinity. *Giga Science* **6**, gix017 (2017).

14. Liu, Z. et al. DeepSeqPan, a novel deep convolutional neural network model for pan-specific class I HLA-peptide binding affinity prediction. *Scientific reports* **9**, 1-10 (2019).

15. Wu, J. et al. DeepHLApan: a deep learning approach for neoantigen prediction considering both HLA-peptide binding and immunogenicity. *Frontiers in immunology* **10**, 2559 (2019).

16. Venkatesh, G., Grover, A., Srinivasaraghavan, G. & Rao, S. MHCAttnNet: predicting MHC-peptide bindings for MHC alleles classes I and II using an attention-based deep neural model. *Bioinformatics* **36**, i399-i406 (2020).

17. Ye, Y. et al. MATHLA: a robust framework for HLA-peptide binding prediction integrating bidirectional LSTM and multiple head attention mechanism. *BMC bioinformatics* **22**, 1-12 (2021).

18. Gfeller, D. et al. The length distribution and multiple specificity of naturally presented HLA-I ligands. *The Journal of Immunology* **201**, 3705-3716 (2018).

19. Reynisson, B., Alvarez, B., Paul, S., Peters, B. & Nielsen, M. NetMHCpan-4.1 and NetMHCIIpan-4.0: improved predictions of MHC antigen presentation by concurrent motif deconvolution and integration of MS MHC eluted ligand data. *Nucleic acids research* **48**, W449-W454 (2020).

20. Rammensee, H.-G., Bachmann, J., Emmerich, N.P.N., Bachor, O.A. & Stevanović, S. SYFPEITHI: database for MHC ligands and peptide motifs. *Immunogenetics* **50**, 213-219 (1999).

21. Kim, Y., Sidney, J., Pinilla, C., Sette, A. & Peters, B. Derivation of an amino acid similarity matrix for peptide: MHC binding and its application as a Bayesian prior. *BMC bioinformatics* **10**, 1-11 (2009).

22. Rasmussen, M. et al. Pan-specific prediction of peptide–MHC class I complex stability, a correlate of T cell immunogenicity. *The Journal of Immunology* **197**, 1517-1524 (2016).

23. Hu, Y. et al. ACME: pan-specific peptide–MHC class I binding prediction through attention-based deep neural networks. *Bioinformatics* **35**, 4946-4954 (2019).

24. Phloyphisut, P., Pornputtapong, N., Sriswasdi, S. & Chuangsuwanich, E. MHCSeqNet: a deep neural network model for universal MHC binding prediction. *BMC bioinformatics* **20**, 1-10 (2019).

25. Collobert, R. et al. Natural language processing (almost) from scratch. *Journal of machine learning research* **12**, 2493− 2537 (2011).

26. Mikolov, T., Chen, K., Corrado, G. & Dean, J. Efficient estimation of word representations in vector space. *arXiv preprint arXiv:1301.3781* (2013).

27. Jensen, K.K. et al. Improved methods for predicting peptide binding affinity to MHC class II molecules. *Immunology* **154**, 394-406 (2018).

28. Dai, Z. et al. Machine learning optimization of peptides for presentation by class II MHCs. *Bioinformatics* **37**, 3160-3167 (2021).

29. Rappazzo, C.G., Huisman, B.D. & Birnbaum, M.E. Repertoire-scale determination of class II MHC peptide binding via yeast display improves antigen prediction. *Nature communications* **11**, 1-14 (2020).

30. Chen, B. et al. Predicting HLA class II antigen presentation through integrated deep learning. *Nature biotechnology* **37**, 1332-1343 (2019).

31. Racle, J. et al. Robust prediction of HLA class II epitopes by deep motif deconvolution of immunopeptidomes. *Nature biotechnology* **37**, 1283-1286 (2019).

32. Henikoff, S. & Henikoff, J.G. Performance evaluation of amino acid substitution matrices. *Proteins: Structure, Function, and Bioinformatics* **17**, 49-61 (1993).

33. UniProt: the universal protein knowledgebase in 2021. *Nucleic Acids Research* **49**, D480-D489 (2021).

34. Cho, K. et al. Learning phrase representations using RNN encoder-decoder for statistical machine translation. *arXiv preprint arXiv:1406.1078* (2014).
35. Vaswani, A. et al. in Advances in neural information processing systems 5998-6008 (2017).
36. Rist, M.J. et al. HLA peptide length preferences control CD8+ T cell responses. *The Journal of Immunology* **191**, 561-571 (2013).
37. Maenaka, K. et al. Nonstandard peptide binding revealed by crystal structures of HLA-B* 5101 complexed with HIV immunodominant epitopes. *The Journal of Immunology* **165**, 3260-3267 (2000).
38. Sarkizova, S. et al. A large peptidome dataset improves HLA class I epitope prediction across most of the human population. *Nature biotechnology* **38**, 199-209 (2020).
39. Zhang, H., Lund, O. & Nielsen, M. The PickPocket method for predicting binding specificities for receptors based on receptor pocket similarities: application to MHC-peptide binding. *Bioinformatics* **25**, 1293-1299 (2009).
40. Mei, S. et al. A comprehensive review and performance evaluation of bioinformatics tools for HLA class I peptide-binding prediction. *Briefings in bioinformatics* **21**, 1119-1135 (2020).
41. O'Donnell, T.J. et al. MHCflurry: open-source class I MHC binding affinity prediction. *Cell systems* **7**, 129-132. e124 (2018).
42. O'Donnell, T.J., Rubinsteyn, A. & Laserson, U. MHCflurry 2.0: Improved pan-allele prediction of MHC class I-presented peptides by incorporating antigen processing. *Cell systems* **11**, 42-48. e47 (2020).
43. Kalaora, S. et al. Use of HLA peptidomics and whole exome sequencing to identify human immunogenic neo-antigens. *Oncotarget* **7**, 5110 (2016).
44. Kalaora, S. et al. Combined analysis of antigen presentation and T-cell recognition reveals restricted immune responses in melanoma. *Cancer discovery* **8**, 1366-1375 (2018).
45. Bassani-Sternberg, M. et al. Direct identification of clinically relevant neoepitopes presented on native human melanoma tissue by mass spectrometry. *Nature communications* **7**, 1-16 (2016).
46. Wang, P. et al. Peptide binding predictions for HLA DR, DP and DQ molecules. *BMC bioinformatics* **11**, 1-12 (2010).
47. Nielsen, M., Lundegaard, C. & Lund, O. Prediction of MHC class II binding affinity using SMM-align, a novel stabilization matrix alignment method. *BMC bioinformatics* **8**, 1-12 (2007).
48. Sturniolo, T. et al. Generation of tissue-specific and promiscuous HLA ligand databases using DNA microarrays and virtual HLA class II matrices. *Nature biotechnology* **17**, 555-561 (1999).
49. Sidney, J. et al. Quantitative peptide binding motifs for 19 human and mouse MHC class I molecules derived using positional scanning combinatorial peptide libraries. *Immunome research* **4**, 1-14 (2008).
50. Vita, R. et al. The immune epitope database (IEDB): 2018 update. *Nucleic acids research* **47**, D339-D343 (2019).
51. Vita, R. et al. The immune epitope database (IEDB) 3.0. *Nucleic acids research* **43**, D405-D412 (2015).
52. Alvarez, B. et al. NNAlign_MA; MHC peptidome deconvolution for accurate MHC binding motif characterization and improved T-cell epitope predictions. *Molecular & Cellular Proteomics* **18**, 2459-2477 (2019).
53. Nielsen, M. et al. NetMHCpan, a method for quantitative predictions of peptide binding to any HLA-A and-B locus protein of known sequence. *PloS one* **2**, e796 (2007).
54. Kawashima, S. & Kanehisa, M. AAindex: amino acid index database. *Nucleic acids research* **28**, 374-374 (2000).
55. Tan, C., Wang, T., Yang, W. & Deng, L. PredPSD: a gradient tree boosting approach for single-stranded and double-stranded DNA binding protein prediction. *Molecules* **25**, 98 (2019).